\documentclass[journal]{IEEEtran}
\usepackage{amsmath,amssymb, amsfonts}
\usepackage{amsthm}
\usepackage{array}
\usepackage[caption=false,font=normalsize,labelfont=sf,textfont=sf]{subfig}
\usepackage[linesnumbered,ruled,vlined]{algorithm2e}
\usepackage{textcomp}
\usepackage{stfloats}
\usepackage{url}
\usepackage{verbatim}
\usepackage{graphicx}
\usepackage{cite}
\usepackage{xcolor}
\usepackage{tabularray}

\usepackage{pbox,adjustbox}
\usepackage{booktabs}      
\usepackage{makecell}      

\usepackage{booktabs,siunitx}
\usepackage[justification=raggedright,singlelinecheck=false,font=footnotesize]{caption}

\usepackage{pifont}

\usepackage[table]{xcolor}
\usepackage{etoolbox}
\makeatletter
\patchcmd{\@makecaption}
  {\scshape}
  {}
  {}
  {}
\makeatletter
\patchcmd{\@makecaption}
  {\\}
  {.\ }
  {}
  {}
\makeatother

\allowdisplaybreaks
\usepackage{enumitem}

\SetKwInput{KwConstants}{Constants}
\SetKwInput{KwOutput}{Output}
\SetKwInput{KwInput}{Input}
\SetKwFor{ClientFor}{each client}{in parallel:}{}
\SetKwFor{SubnetFor}{each subnet}{in parallel:}{}
\SetKwFor{AggrFor}{all sampled clients}{aggregate:}{}
\newlist{myenumi}{description}{10}
\setlist[myenumi]{labelindent=\parindent, leftmargin=*, label=(\roman*), align=left}
\setlist[myenumi]{leftmargin=0pt}
\definecolor{lightgreen}{RGB}{181, 199, 147} 

\begin{document}

\title{Federated Foundation Models in Harsh Wireless Environments: Prospects, Challenges, and Future Directions}

\author{Evan Chen,
~Seyyedali Hosseinalipour, 
~Christopher G. Brinton,
~David J. Love

\thanks{Evan Chen, Christopher G. Brinton, and David J. Love are with the Elmore Family School of Electrical and Computer Engineering, Purdue University, West Lafayette, IN, 47907, USA. E-mail: \{chen4388,cgb,djlove\}@purdue.edu.}
\thanks{S. Hosseinalipour is with  University at Buffalo, SUNY, 12 Capen Hall
Buffalo, NY, 14260, USA. E-mail: alipour@buffalo.edu.}
}

\maketitle

\begin{abstract}
Foundation models (FMs) have shown remarkable capabilities in generalized intelligence, multimodal understanding, and adaptive learning across a wide range of domains. However, their deployment in \textit{harsh or austere environments} -- characterized by intermittent connectivity, limited computation, noisy data, and dynamically changing network topologies -- remains an open challenge. Existing distributed learning methods such as federated learning (FL) struggle to adapt in such settings due to their reliance on stable infrastructure, synchronized updates, and resource-intensive training. In this work, we explore the potential of Federated Foundation Models (FFMs) as a promising paradigm to address these limitations. By integrating the scalability and generalization power of FMs with novel decentralized, communication-aware FL frameworks, we aim to enable robust, energy-efficient, and adaptive intelligence in extreme and adversarial conditions. We present a detailed breakdown of system-level constraints in harsh environments, and discuss the open research challenges in communication design, model robustness, and energy-efficient personalization for these unique settings.
\end{abstract}


\section{Introduction}
\noindent Machine Learning (ML) has revolutionized numerous domains, from healthcare and smart cities to transportation and environmental monitoring~\cite{jordan2015machine}. In their conventional forms, ML models are trained and deployed to handle a single modality of data (e.g., image, text) and a single downstream task (e.g., classification of a particular dataset). However, intelligent services offered in modern wireless networks are increasingly demanding the simultaneous processing of multiple modalities of data collected at wireless devices, and handling the execution of a variety of downstream tasks. For example, in smart driving cars, image, LiDAR, and sensor readings are jointly leveraged to support tasks such as collision avoidance, traffic sign recognition, and driver assistance.


We have witnessed a similar trend towards multi-task and multi-modality in the ML area. The community is migrating towards ML models capable of conducting multiple ML tasks through the use of various input data modalities. The advent of Large Language Models (LLMs) was a giant leap in this direction, as LLMs are capable of executing multiple language/text-related tasks (e.g., code synthesis, text generation, summarization, etc.). More recently, the need for ML models with larger scopes led to another major innovation in the ML area: \textit{Foundation models (FMs)}. FMs enable the use of simultaneous modalities of data (through dedicated encoders), such as text, audio, image, and video, to execute multiple downstream tasks (through dedicated task heads), such as audio generation, translation, image generation, video generation, and text refinement. Further, through their unified architecture, FMs can enable knowledge sharing across multiple modalities, transfer knowledge across tasks, and swiftly adapt to unseen data through lightweight Parameter Efficient Fine-Tuning (PEFT). Popular PEFT techniques include adapters, Low-Rank Adaptations (LoRA), and prompt tuning~\cite{chang2024survey}. Notably, their strong zero/few-shot learning capabilities have made PEFT particularly appealing for \textit{on-device fine-tuning and inference}, especially when considering resource-constrained devices. 

\begin{figure}
    \centering
    \includegraphics[width=0.95\linewidth]{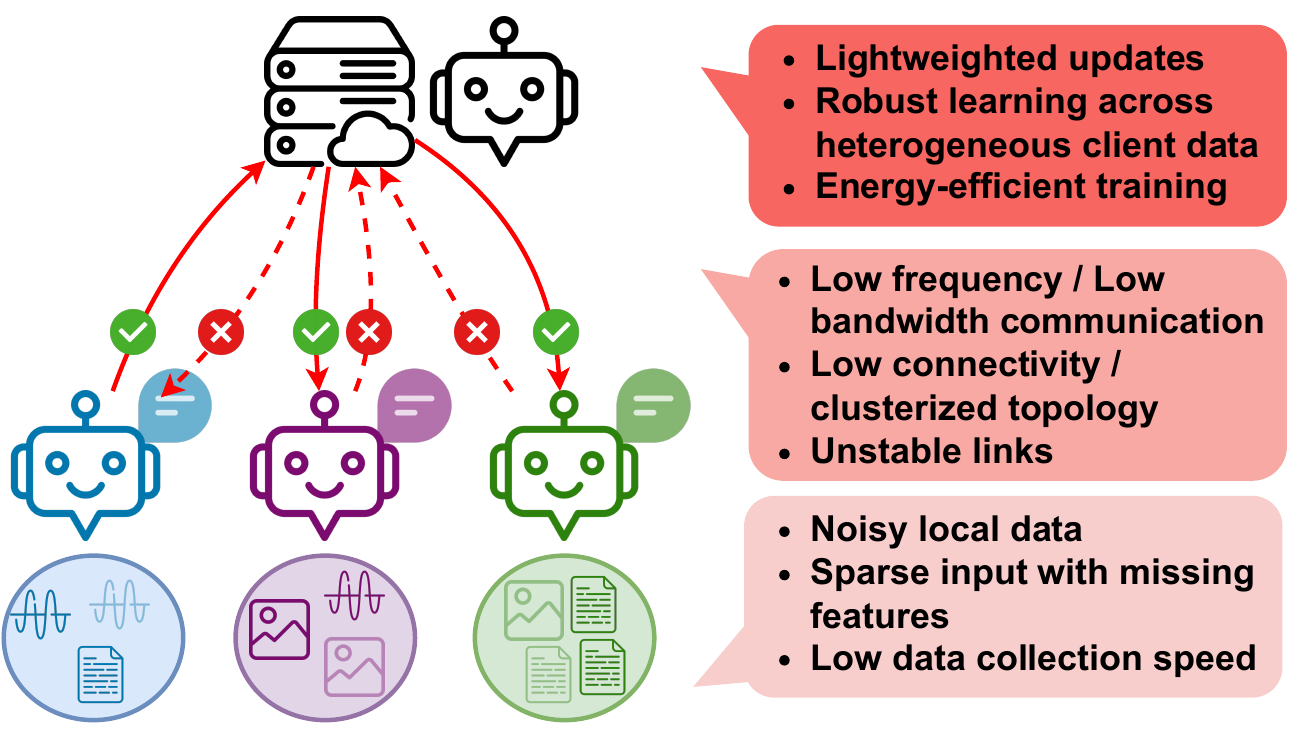}
    \caption{Multiple levels of deployment challenges arise when aiming to conduct Federated Learning (FL) in harsh environments, including unreliable communication layers and noisy, incomplete local data for training which create bottlenecks for server-side coordination.}
    \label{fig:fl_challenge}
\end{figure}


Yet, due to their nascent nature, FMs have not yet been deployed across networks with unusual or extreme requirements, which the communication systems community collectively refers to as \textit{harsh or austere wireless networks/environments}. In particular, \textit{in this work, for the first time in the literature, we shed light on the deployment of FMs in harsh environments.} Harsh environments (encountered in e.g., undersea exploration, underground mines, polar regions) often yield data that are sparse, corrupted by sensor failures/miscalibration, or biased toward narrow operating regimes. ML models trained solely on these local datasets will thus tend to converge to suboptimal solutions and perform poorly when exposed to changing scenarios. This calls for unlocking the access of FMs to distributed data collected across multiple harsh environment regions during their fine-tuning/training stage. Given well-known challenges in centralizing large volumes of data collected across these regions, doing so requires fine-tuning/training according to distributed learning techniques. 


Federated Learning (FL)~\cite{kairouz2021advances} is a popular technique for distributed ML which allows multiple devices to collaboratively train a shared global model without exchanging raw data, thereby preserving privacy and reducing communication overhead. Consequently, it is compelling to consider leveraging FL-driven fine-tuning/training of FMs, referred to as \textit{Federated Foundation Models (FFMs)}~\cite{zhuang2023foundation}, in harsh environments. However, there are only a few existing works on FFMs~\cite{ren2025advances}, and the literature is yet to explore them in harsh environments, motivating this study.
In particular, the operational realities of such environments, ranging from intermittent connectivity and physical hazards to extreme resource constraints and rapidly shifting data distributions (depicted in Fig.~\ref{fig:fl_challenge}), demand a careful rethinking of how FFMs are implemented in these environments~\cite{el2019case,feng2023fedmultimodal}.

\textbf{Summary of contributions.} In this work, our contributions are as follows:
\begin{itemize}
\item We highlight the advantages of FMs for providing intelligence in harsh environments, outlining their capabilities for multimodal perception, rapid adaptability, task versatility, and robustness to noise and distribution shifts.
\item We identify the unique challenges that arise when deploying FFMs in harsh environments, including the impacts of fragile and failure-prone devices, asynchronous and stale model updates, and the difficulty of maintaining or replacing nodes in inaccessible locations.
\item We outline open research directions for realizing FFMs in harsh environments, including multimodal alignment, drift mitigation, robustness to unreliable communication and corrupted data, and energy-efficient operation.
\end{itemize}

\begin{figure}
    \centering
    \includegraphics[width=0.95\linewidth]{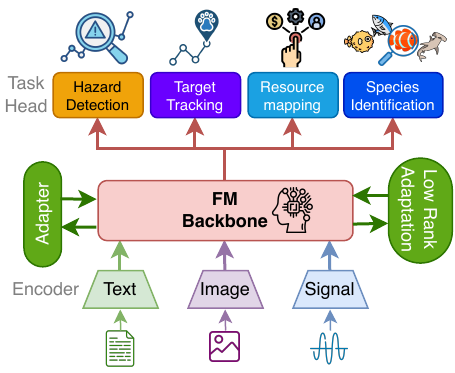}
    \caption{Illustration of an FM architecture. Multimodal inputs are processed by a set of encoders and passed to a  shared backbone. Parameter-efficient tuning methods, including adapters and low-rank adaptation, enable lightweight specialization. The unified representation produced by the backbone are used to support diverse downstream tasks.}
    \vspace{-0.1in}
    \label{fig:fm_architecture}
\end{figure}

\begin{figure*}
    \centering
    \includegraphics[width=0.95\linewidth]{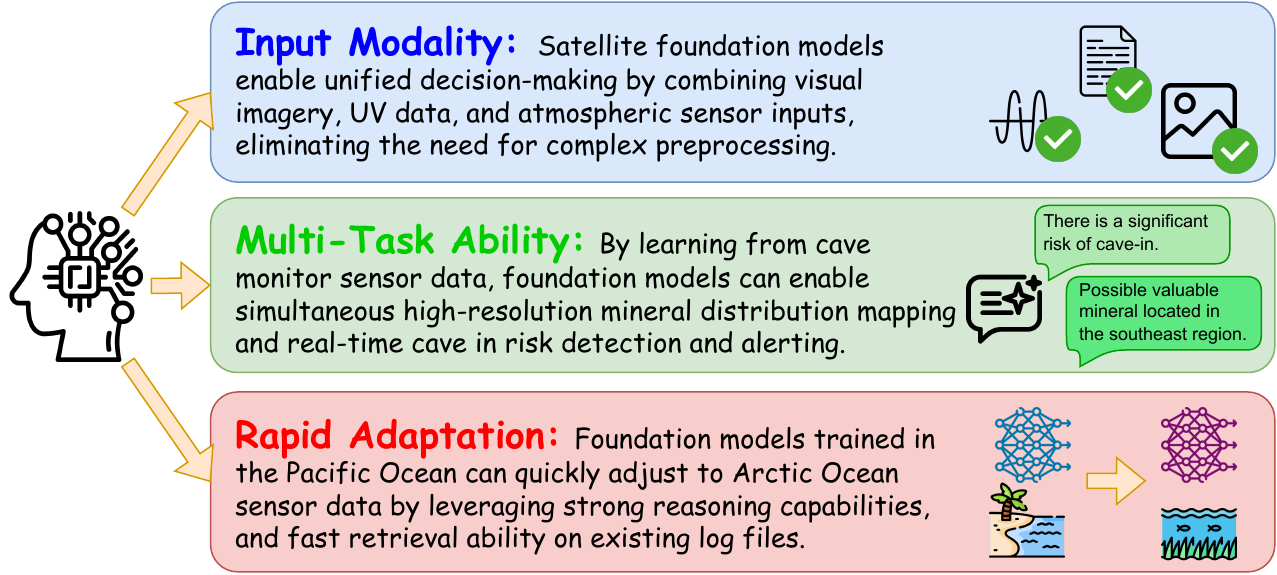}
    \caption{Foundation Models (FMs) have the potential to enable robust, multimodal, and adaptable intelligence across heterogeneous harsh environments, with resilience to noise and distribution shifts. Combined with Federated Learning (FL), FFMs empower edge devices to collaboratively perceive, reason, and act under extreme physical and communication constraints.}
    \label{fig:FM_capability}
\end{figure*}

\section{Key Concepts of FFMs}
\label{sec:motivations}
\subsection{Structure of FMs}

FMs are large-scale ML architectures pre-trained on massive, diverse corpora to learn general-purpose representations across modalities and tasks. As illustrated in Fig.~\ref{fig:fm_architecture}, a typical FM can be viewed as three components:

\textbf{1. Encoder:} 
Modality-specific encoders (e.g., vision, audio, text) transform raw inputs into latent embeddings. Common choices include transformer encoders (e.g., Vision Transformers for images and BERT/T5-style encoders for text) and Convolutional Neural Network (CNN) or hybrid stacks that combine CNNs and transformers for audio and sensor data. 

\textbf{2. Backbone:} 
The encoder embeddings are processed by a multi-layer transformer backbone that enables cross-modality reasoning. There are mainly two widely used designs:
i) Fusion-based multimodal transformer: In this backbone, embeddings from all encoders are fused within a shared transformer (e.g., Flamingo-style language backbones augmented with gated cross-attention; BLIP-2 with a Q-Former bridge). The backbone is explicitly trained to integrate signals across modalities.
ii) Decoder-only LLM backbone with adapters: In this backbone, outputs from non-text encoders are aligned to the language token space via adapters, after which a decoder-only LLM (e.g., GPT/LLaMA/Mixtral backbones) performs autoregressive reasoning over a unified token sequence. In this backbone, multimodality is achieved by alignment rather than fusion.

\textbf{3. Task Heads:} 
Given the shared representation from the backbone, lightweight heads (e.g., linear layers or shallow MLPs) implement task-specific outputs such as classification, regression, or retrieval. 

\subsection{Advantages of FMs}

There are mainly three practical advantages of FMs naturally emerge from its structure:

\textbf{1. Flexible input modality:} The modularized encoder design in FMs allows independent process of each modality. This structure makes FMs particularly well-suited for environments with time-varying or incomplete modality availability. For instance, unmanned underwater vehicles (UUVs) may collect sonar, acoustic, infrared, or optical data, but not all sensors may be active at every time step~\cite{li2024multimodal}. Rather than relying on fixed input configurations, the model can seamlessly operate with partial observations, using masking or cross-attention mechanisms to dynamically fuse available embeddings. Moreover, new encoders can be swapped in or updated without altering the backbone, enabling rapid adaptation to evolving sensing setups.

\textbf{2. Robustness to heterogeneity and noise:} 
FMs demonstrate strong robustness to distributional shifts and sensor noise, which is largely attributed to their causal reasoning nature. Whether dealing with random ambient disturbances in acoustic channels, seasonal changes in forest imagery, or transient interference in wireless networks, the FM backbone learns to emphasize stable semantic patterns over surface-level noise. The semantic understanding capability of FM backbones can implicitly regularize against corrupted or missing data segments. 

\textbf{3. Versatility across multiple tasks:} The modular structure also extends to the task layer. Since the backbone is responsible for extracting general-purpose, high-level representations, the downstream heads required for specific tasks can remain shallow and lightweight. This separation of understanding from execution enables efficient multitasking: a single FM can support numerous objectives without retraining its backbone. As a result, FMs not only generalize across modalities and conditions but also scale gracefully across tasks, supporting diverse downstream applications with minimal customization.

\subsection{Training/Fine-Tuning of FFMs}
The training/fine-tuning of FFMs are governed by iterative FL protocols, in which clients perform local updates using their private data and periodically synchronize with other participants through communication rounds. This collaborative training process enables the global model to learn from diverse, decentralized datasets without raw data ever leaving the local devices. Due to the scale and modular structure of FMs, this procedure must account for both communication efficiency and computational feasibility. As such, the training framework can be naturally decomposed into three tightly coupled aspects: \textit{1) local training}, which determines how each client updates its model under resource constraints; \textit{2) network architecture}, which defines how information is exchanged across clients through a communication system; and \textit{3) aggregation method}, which governs how local updates are integrated into the global model.

\textbf{1. Local Training:} 
At the beginning of training, each client may either initialize its own model parameters or download a pre-trained model from the server. Local training is then performed using each client's private dataset, which reflects the unique sensing conditions, modalities, and environmental variations of that particular device. Given the large number of parameters in FM backbones, full-model training is often impractical. Fortunately, the modular structure of FMs, with decoupled encoders, backbone, and task heads, allows partial updates that reduce local computational and memory costs. As a result, PEFT techniques are commonly employed. For transformer-based encoders and backbones, methods such as \textit{Adapters} (which insert small trainable bottleneck layers within attention modules) or \textit{LoRA} (which performs low-rank adaptation of attention weights) enable adaptation to local data distributions with minimal overhead. For task heads, freezing the backbone and updating only the output layers can provide client-specific specialization without overfitting to multimodal variations. Collectively, these techniques allow local updates to be both computationally efficient and semantically expressive.

\textbf{2. Network Design:}
Once local training completes, clients must exchange information to obtain information from other clients. The structure of the communication network plays a crucial role in determining both efficiency and convergence. Three primary designs are commonly considered. \textit{i) Centralized networks:} A single server coordinates the learning process by collecting model updates from all clients and broadcasting the aggregated model back. This setup is often the most effective in terms of convergence guarantees and model performance, but it suffers from scalability issues due to the communication bottleneck at the server, particularly when handling thousands of clients or large-scale FM updates. \textit{ii) Decentralized networks:} To address this, decentralized topologies remove the central coordinator entirely. Clients exchange updates with their peers via a predefined graph (e.g., ring, random geometric). This structure improves scalability and robustness to node failure, especially in bandwidth-constrained or intermittently connected environments. However, convergence may degrade significantly, especially when the connectivity graph is sparse or unbalanced. \textit{iii) Hierarchical networks:} As a middle ground, hierarchical networks organize clients into multi-tier structures (e.g., client - edge - cloud). Clients first send updates to local aggregators (e.g., fog nodes), which then communicate with higher-level servers. This design reduces per-node communication load while preserving relatively strong convergence properties, making it well-suited for large-scale deployments.

\textbf{3. Aggregation Methods:} 
The modularized structure of FMs allows two choices for model aggregation. \textit{i) Weighted Average aggregation:} When all clients share an identical FM architecture (i.e., same encoders, backbone, and heads), standard Federated Averaging (FedAvg) can be applied. Each client transmits its model or update, and the server performs weighted averaging across layers to synthesize a global model. This method is simple and well-understood, but becomes limiting in cases where clients exhibit modality- or task-specific variation. \textit{ii) Mixture-of-Experts (MoE) aggregation:} To support more diverse deployments, recent work proposes splitting FM backbones into multiple modality-specific and task-specific expert modules. The server maintains a global pool (the union of all expert modules), where each client holds a subset of expert modules related to the local environment it is located in. This approach enables heterogeneous clients to participate meaningfully in training while maintaining communication efficiency. Furthermore, MoE-style aggregation aligns naturally with modular FMs and supports personalization and efficient scaling across tasks and environments.

\begin{figure}
    \centering
    \includegraphics[width=0.95\linewidth]{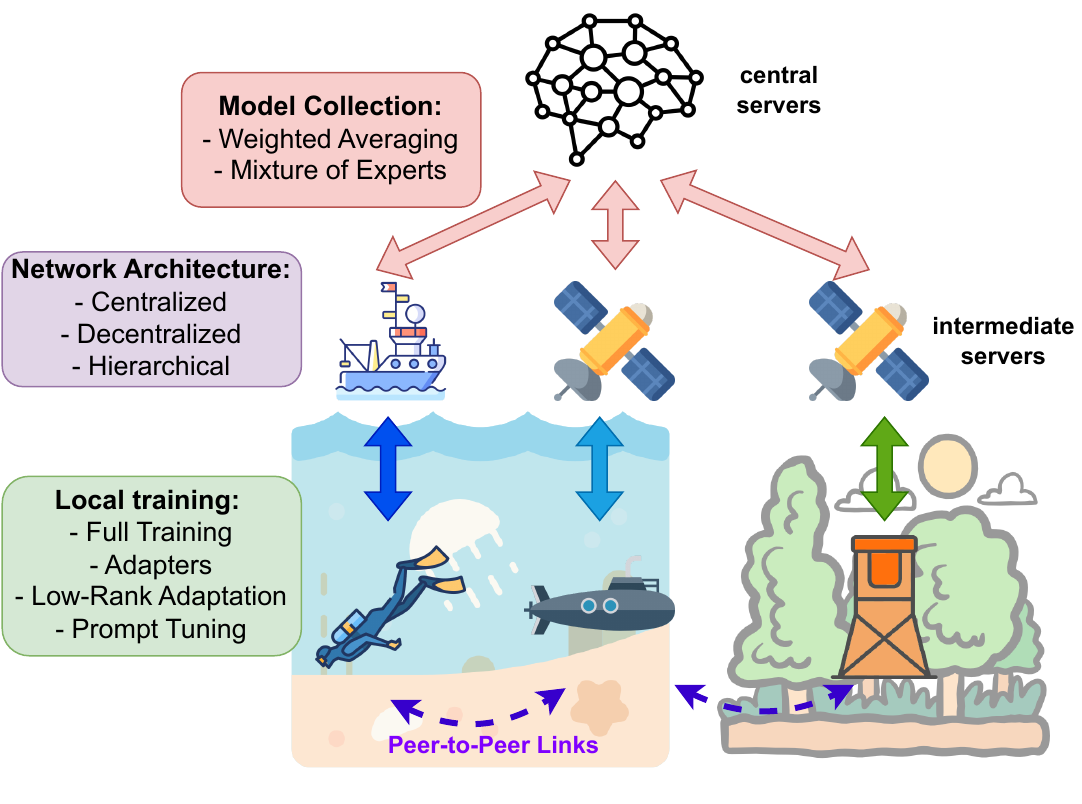}
    \caption{Overview of FFM learning architectures. Edge devices perform local training and communicate updates through centralized, decentralized (peer-to-peer), or hierarchical topologies. Model collection leverages strategies such as weighted averaging and mixture of experts for training optimization. }
    \label{fig:ffm_procedure}
    \vspace{-0.1in}
\end{figure}

$\star$ This cycle of local update and aggregation is repeated for multiple communication rounds as shown in Figure~\ref{fig:ffm_procedure}, allowing the FFM to progressively integrate knowledge from diverse environments while maintaining adaptability to local conditions.

\section{FFMs in Harsh Environments: Motivation and Core Challenges}

\subsection{Definition of Harsh Environments}
In harsh environments, the foundational assumptions behind current distributed learning infrastructure (reliable communication, abundant energy, and stable sensing) often fail to hold. These settings impose compounded physical and network-level constraints: extreme pressure, temperature, salinity, vibration, and structural instability can directly impair sensing and computation, while high latency, intermittent links, and limited or asymmetric bandwidth degrade communication reliability. We highlight three core characteristics that define the operational challenges of harsh environments:

\textbf{Necessity of Learning-Based Control and Decision-Making:}  
Traditional rule-based control or reactive systems lack the flexibility to handle the unpredictable and non-stationary conditions common in harsh environments. Devices must go beyond fixed logic and incorporate learning capabilities to detect patterns, adapt to novel inputs, and make decisions under uncertainty. However, learning in these settings is inherently challenging: data is fragmented across distributed nodes, connectivity is unreliable, and centralized data collection is often infeasible. This raises the need for learning frameworks that can operate effectively under partial observability and limited supervision.

\textbf{Need for Collaborative, On-Device Intelligence:}  
The sparsity and fragility of communication links makes frequent synchronization or coordination with a central server infeasible, requiring devices to operate with high levels of autonomy and local awareness. In this setting, devices must learn from their own sensor data, adapt in real time, and share insights opportunistically with peers or intermediate edge nodes. These constraints are compounded by high heterogeneity across devices, both in sensing modalities and in the nature of local observations. As a result, achieving generalizable behavior across a distributed system becomes increasingly difficult without collaborative, resource-aware intelligence distributed across the network.

\textbf{Breakdown of Conventional Communication:}  
Communication at the physical layer is often unreliable or entirely unavailable in harsh environments. Lower-frequency bands such as vintage VHF offer stronger propagation and penetration in obstructed or subsurface conditions, but suffer from low capacity and are vulnerable to interference. In more extreme scenarios (e.g. deep underwater or underground) electromagnetic signaling becomes infeasible, requiring the use of alternative modalities like acoustic or pressure-wave communication. While such methods are widely adopted in industrial practice (e.g., through rock, fluid, or pipeline media), they remain poorly characterized in academic communication theory. Fundamental signal models, capacity limits, and protocol designs must be revisited to accommodate these unconventional media.

\subsection{Examples of Harsh Environments}
To ground these concepts, we highlight four representative settings that impose distinct physical and link-layer characteristics:

 \textbf{1. Underground Mines} suffer from non-line-of-sight propagation, metallic and geological interference, and dynamically evolving tunnel geometries that change signal paths as excavation proceeds. Hazards like collapses or gas leaks can abruptly sever communication, demanding local autonomy and safety-aware learning.
 
 \textbf{2. Underwater Acoustic Networks} face low bandwidth, long and variable propagation delays, Doppler effects, and nonlinear refraction due to stratification. These challenges require models that are robust to delayed or distorted feedback and can operate reliably on sparse updates.
 
 \textbf{3. Forest Deployments} must contend with dense vegetation and variable weather, both of which attenuate and scatter signals. Communication often depends on opportunistic encounters (e.g. drones or wildlife-tagged sensors coming into proximity) which favors asynchronous, peer-to-peer model exchange.
 
 \textbf{4. Satellite Constellations} experience highly dynamic topology due to orbital motion, fast handovers, and intermittent ground station access. Power constraints, onboard compute limitations, and strict latency windows further challenge traditional learning pipelines.

$\star$ These examples characterize the operational backdrop for learning systems in harsh environments. Section~\ref{ssec:ffm_challenges} translates these physical and link-layer realities into concrete challenges for training and inference, while Figure~\ref{fig:harsh_environments} summarizes the mapping from physical constraints to network effects.

\subsection{Why FFMs are needed in Harsh Environments}

The unique structure and training procedure of FFMs make them especially well-suited for deployment in harsh environments, where traditional ML models struggle due to sensor unreliability, communication bottlenecks, and distribution shift. In such settings, each individual device operates under localized conditions, observes only a narrow slice of the world, and may encounter degraded or missing modalities. FFMs address these challenges through three core capabilities:

\textbf{1. Robustness to local degradation and partial observability:}
Harsh environments often introduce sensor noise or missing modalities due to energy constraints or adverse conditions. By leveraging the modular encoder design of FMs, clients can flexibly process whichever modalities are available at a given time. Moreover, the causal reasoning capability of FM backbones help stabilize inference under non-stationary inputs. However, local training alone is not enough—if a model is trained solely on corrupted or incomplete data from one device, it may overfit to local data distributions. Hence, FFMs plays a key role here by allowing models to pool knowledge across devices, smoothing out the impact of sensor-level degradation and reinforcing common patterns observed across the network.

\textbf{2. Adaptation to dynamic and non-i.i.d. environments:}
In extreme settings, the data distribution varies not only across modalities but also significantly across clients, driven by mobility, terrain shifts, seasonal variation, or mission-specific objectives. In such cases, the heterogeneity can be so pronounced that local datasets across clients may become entirely disjoint in content or modality coverage. Despite this fragmentation, the collective knowledge distributed across clients is essential for constructing a general-purpose and robust FM. While FMs are highly capable of local adaptation, adaptation without any form of cross-client coordination leads to representational drift, where each client model overfits to its local environment and fails to learn transferable reasoning capabilities. FFMs address this issue by periodically aggregating model updates across the network, allowing locally adapted models to realign toward a shared semantic space. This process preserves each client’s specialization while reinforcing common structure across the network. Maintaining this balance is critical for enabling devices to collaborate effectively and share insights.

\textbf{3. Scalability and resilience in resource-constrained networks:}
Communication infrastructure in harsh environments is often sparse, delayed, or unreliable, due to low-bandwidth links, physical obstructions, and energy limitations. These constraints make frequent, full-model synchronization infeasible and create bottlenecks for centralized coordination. To address this, FFMs leverage the modular structure of foundation models to enable partial aggregation: instead of transmitting the entire model, clients can selectively share updated components based on their local data distribution. Moreover, to further reduce communication load and improve robustness, FFMs can adopt hierarchical or decentralized architectures that eliminate reliance on a single server. In these networks, clients exchange updates with local aggregators or direct peers, allowing scalable and distributed coordination even under limited connectivity. This combination of structured aggregation and topology-aware communication strategies enables FFMs to operate efficiently and reliably in bandwidth-constrained and intermittently connected environments.

$\star$ FFMs integrate three essential capabilities that are fundamental for reliable operation in extreme conditions: robustness to degradation, adaptability to heterogeneity, and scalability under constrained communication. Without federation, isolated FMs would risk becoming overfitted and inconsistent. With federation, FFMs are capable of learning from and reacting to local challenges while maintaining global coherence.

\subsection{Challenges of FFMs in Harsh Environments}
\label{ssec:ffm_challenges}
\begin{figure*}
    \centering
    \includegraphics[width=0.95\linewidth]{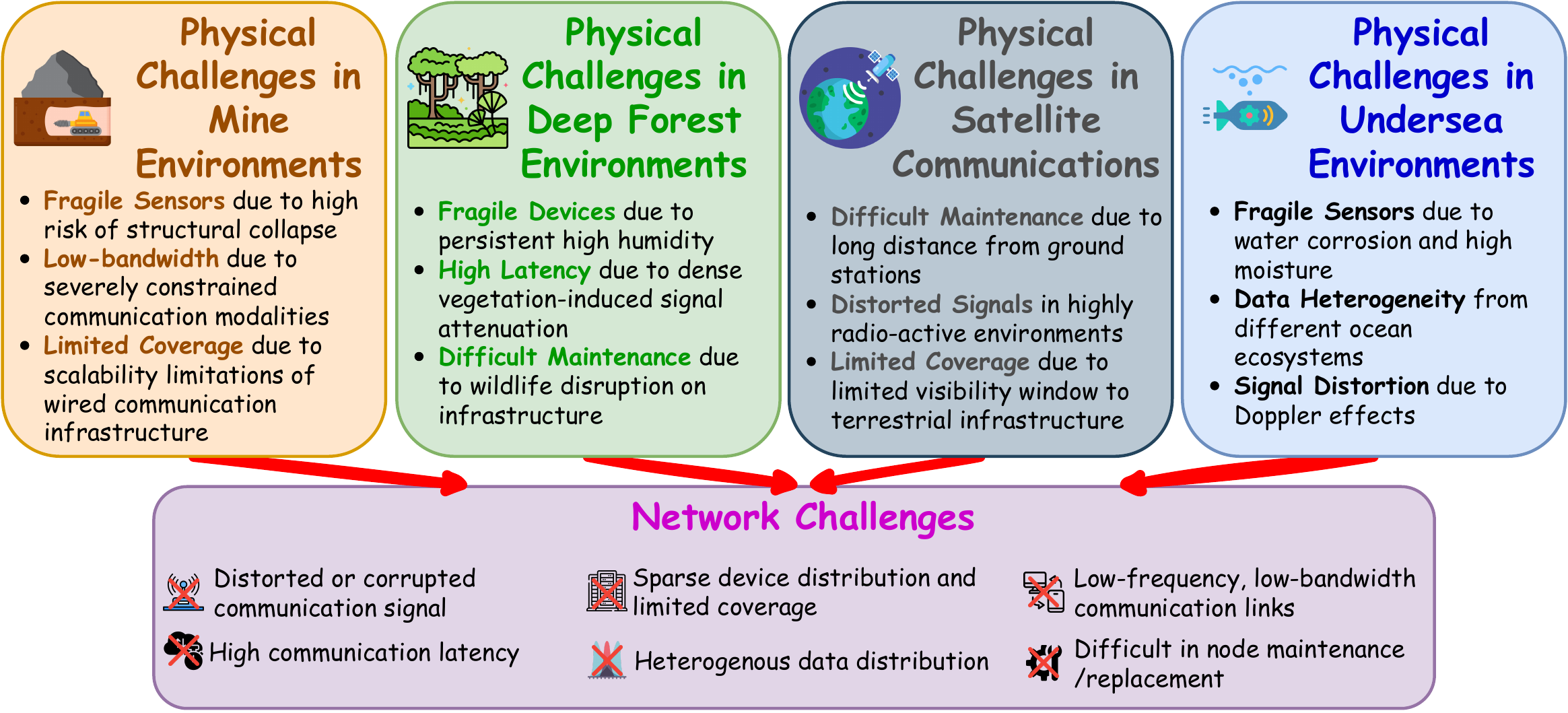}
    \caption{Physical and network challenges for Federated Learning (FL) in harsh environments. Each category of harsh environment presents distinct physical constraints such as structural collapse risk, signal attenuation, and extreme humidity. These physical limitations give rise to a shared set of network-level challenges that hinder reliable FL deployment. The issues must be addressed to design robust, energy-efficient, and adaptive FL systems across diverse hostile environments.}
    \label{fig:harsh_environments}
\end{figure*}

The physical constraints of harsh environments can be translated into specific network-level challenges (depicted in Figure~\ref{fig:harsh_environments}), which in turn impact the design of learning systems. In this section, we identify several network challenges that emerge from harsh environments and explain how they affect FFM training and inference.

\textbf{1. Fragility of Deployed Devices and Sensors:}  
Harsh environments impose mechanical, environmental, and chemical stress on deployed equipment. In underground mines, for instance, heavy vibrations, rockfalls, and high dust concentration frequently damage communication relays and sensors~\cite{yarkan2009underground}. Similarly, underwater deployments face corrosion from saline water, high-pressure fluctuations, and biofouling~\cite{fattah2020survey}, while forest environments expose devices to extreme humidity, rainfall, and wildlife interference. Frequent failures and intermittent sensing produce missing, partial, or corrupted updates. FFMs must tolerate update dropouts, filter unreliable signals, and bound the impact of noisy or adversarial contributions. Practical mechanisms include failure-aware aggregation, robust objectives (e.g., clipping or trimmed means), and confidence-weighted fusion that down-weights uncertain updates.

\textbf{2. Asynchronous and Stale Updates in Harsh Networks:}  
Many harsh environments are characterized by significant delays in data transmission, primarily due to physical constraints of the communication medium. For instance, underwater acoustic links suffer from inherently high latency due to the slow propagation speed of sound in water. Satellite-to-ground communication is limited by orbital dynamics and long round-trip times, particularly in deep space scenarios~\cite{kodheli2020satellite}. Similarly, underground and forested environments rely on multi-hop relays and suffer from signal attenuation caused by physical obstructions, introducing additional delay and jitter. These factors lead to irregular or unpredictable contact windows, making tightly synchronized communication difficult or impossible. As a result, updates exchanged between devices are often stale or inconsistent, which degrades the stability and convergence of any distributed learning or control process operating over such networks.

\textbf{3. Sparse and Bandwidth-Constrained Communication:}  
There are mainly two issues in network infrastructure that severely deteriorates the training performance:
\textit{i) Sparse device deployments} are common in deep forests, where dense vegetation and uneven terrain hinder dense sensor placement, and in underwater and satellite systems, where devices are widely spaced due to physical or operational limitations. This spatial sparsity reduces the frequency of inter-device communication and limits opportunities for collaborative learning. Compounding this challenge, \textit{ii) bandwitdth-constrained and unreliable links} further limits the large-scale model transmission and high frequency updates. For example, underwater acoustic channels offer only low-frequency, narrowband connectivity and are susceptible to multipath distortion and high bit error rates~\cite{fattah2020survey}. Similarly, LEO satellite links provide limited contact windows with tight bandwidth and energy budgets~\cite{han2024cooperative}, while undermine deployments typically rely on low-rate radios. These limitations restrict both the volume and fidelity of transmitted model updates. 

\textbf{4. Intensified Data Heterogeneity and Limited Local Diversity:}  
In harsh environments, data heterogeneity arises from a combination of distributional variability and unstable sensing conditions. Two main factors lead to highly non-i.i.d. data across devices:
\textit{i) Modality and Sensory Mismatch:} Devices in the network may be equipped with different sensing modalities (e.g., acoustic, optical, infrared), resulting in heterogeneous input spaces. Even among devices with the same modality, physical separation and localized environmental factors (e.g., ambient noise, pressure gradients) can cause substantial variation in the collected data. \textit{ii) Temporal drift:} The dynamic nature of harsh environments introduces rapid shifts in data distribution. Weather changes or mission-related transitions can alter local data statistics within hours. Sensor failures or degradation (e.g., corrosion, debris buildup) may further reduce sensing capability, leaving nodes with partial or degraded views. 
As a result, each client not only observes a biased slice of the global data distribution but also lacks the intra-client variety needed for stable local training. These conditions severely limit the effectiveness of traditional FL personalization strategies. 



\textbf{5. Difficulty in Node Maintenance and Replacement:}  
Servicing deployed devices in extreme environments is often infeasible. For example, battery replacement for underwater or satellite nodes involves expensive and infrequent missions, while mine and forest deployments face safety and accessibility issues. As a result, nodes are expected to operate autonomously over extended periods under strict energy budgets. 


$\star$ Together, these network challenges underscore the need for tailored FFM strategies that go beyond traditional FL techniques. In particular, FFMs must support asynchronous, low-overhead, and fault-tolerant learning while preserving alignment across multimodal tasks, even when the underlying network is unstable, fragmented, or evolving.

\section{Open Research Directions}
\label{sec:directions}


In this section, we outline critical open problems and propose directions for future research that could further enhance the robustness, adaptability, and effectiveness of FFMs in extreme operational settings.

\textbf{Multimodal Context Alignment:}  
In harsh environments, sensor noise, modality dropouts, and asynchronous data availability can fragment the embedding space, threatening semantic consistency across the network. We envision \textit{environment-aware alignment mechanisms} that enable encoders to stitch together incomplete and noisy cross-modal signals, preserving coherence under strict communication and computation constraints. To achieve this, we envision \textit{shared latent anchors}, a compact set of globally consistent, environment-conditioned reference embeddings that act as semantic landmarks for aligning local representations across clients. These anchors can be periodically broadcast or collaboratively learned, adapting to each client’s sensing and environmental context. To complement this, \textit{Cross-Client Contrastive Regularization} schemes may allow clients performing related tasks to exchange lightweight prototypes or feature statistics, enabling local contrastive alignment based on prior understanding of client tasks.


\begin{table*}[t]
\caption{ Publicly available datasets collected in harsh environments across diverse sensing modalities and tasks. These datasets span satellite, forest, and underwater settings, offering multimodal inputs such as multispectral images, point clouds, and depth maps. They support a range of downstream tasks including species identification, debris detection, and structural inspection, providing valuable benchmarks for developing and evaluating FFMs under challenging real-world conditions. 
\vspace{-1em}}
\label{table:dataset}
\begin{center}
\resizebox{0.95\textwidth}{!}{
\renewcommand{\arraystretch}{1.7}
\begin{tabular}{|cccc|}
\hline
\rowcolor{blue!20} Dataset & Environment & Modalities & Tasks\\
\hline
\rowcolor{lightgreen!90} FLAIR-HUB & Satellite and UAV & Topographic Information, Satellite Image Time Series(SITS), VHR & Land Cover Segmentation, Crop Type
Segmentation\\
\rowcolor{lightgreen!20} Planted & Satellite & SITS, VHR & Forest plantation/species Identification\\
\rowcolor{lightgreen!90} MADOS & Satellite & Multispectral Satellite Images & Marine debris and oil spill mapping\\
\rowcolor{lightgreen!20} FinnWoodlands & Deep Forest & RGB images, point clouds, depth maps & Instance Segmentation, Panoptic Segmentation, Depth Completion \\
\rowcolor{lightgreen!90} SUIM & Underwater & Image & Underwater Species segmentation \\
\rowcolor{lightgreen!20} LIACI & Underwater & Image & Underwater Ship Inspection\\

\hline
\end{tabular}
}\\
\end{center}
\end{table*}

\textbf{Mitigating Multi-Source Drift:}  
In FFM systems deployed in harsh environments, multiple forms of drift including distributional, embedding, and identifier-level shifts can combine to degrade global model performance and consistency over time. We envision \textit{drift-aware FFM optimization frameworks} that incorporate continual adaptation modules to track and compensate for evolving semantic representations and ID mappings. For example, \textit{embedding replay buffers} can be locally maintained to contrast current embeddings with temporally anchored versions, enabling clients to detect and correct semantic drift in their representations. Complementarily, we envision \textit{parameter-efficient gradient tracking} mechanisms that inject lightweight correction terms into the locally computed stochastic gradients, aligning local updates with globally relevant directions and reducing the effects of complex data distributional drifts across asynchronous clients. 

\textbf{Robustness Tuning Under Missing Communication and Corrupted Data:}  
In harsh environments, robustness must withstand both incomplete data and unreliable communication. Prolonged disconnections can leave clients isolated without updates, while sensor failures or miscalibrations may inject corrupted signals into local training, threatening model reliability.
We envision \textit{communication-aware robustness tuning} mechanisms that adapt learning rates or aggregation weights based on the freshness and reliability of updates, ensuring stable progress and eliminate stale updates even under sparse participation. In parallel, we envision \textit{corruption-resilient training objectives} that downweight or filter unreliable local signals using noise-aware loss functions or confidence-based data selection, preventing quality degradation of data from effecting the reasoning ability of FFMs.

\textbf{Energy-Efficient and Sustainable Operation:}
In harsh environments where power supply is limited or intermittent, FFMs must operate within strict energy budgets for both local computation and communication transmission.  
We envision energy-aware training schedules that dynamically adjust batch sizes, update frequencies, or participation levels based on real-time device power states (e.g., adapting learning intensity when solar-charged batteries are low). Additionally, we envision \textit{client-based FM-driven federation protocol optimization}, where the reasoning ability of FMs is leveraged to adapt communication strategies in real time. For example, FFMs could learn semantic communication schemes to compress messages, select efficient channels under acoustic or radio links, or even optimize scheduling and physical-layer operations across the network. Such protocol-level intelligence enables sustainable and self-maintaining federation without constant third-party supervision, ensuring long-term operation under severe energy constraints.

\textbf{Benchmarking in Realistic Harsh Environments:}  
Current benchmarks~\cite{wang2024fedmeki,kikaki2024detecting,islam2020semantic} (Table~\ref{table:dataset}) advance standardized evaluation for federated multimodal systems but lack key features for harsh-environment scenarios. We envision a jointly designed benchmark that incorporates:  
(i) \textit{Environmental stressors} (e.g., humidity, corrosion, radiation);  
(ii) \textit{Device-level heterogeneity} in compute, memory, and energy constraints; (iii) \textit{Robustness testing} for partial signal loss and multimodal desynchronization; and  
(iv) \textit{New evaluation protocols} with metrics such as energy per update, adaptation latency, and resilience under prolonged disconnection.  
Such benchmarks are essential to guide the development of FFMs for real-world deployment in extreme settings.


\section{Conclusion}
We advocate for the development of Federated Foundation Models (FFMs) as a promising paradigm for enabling intelligent systems in harsh environments. By combining the generalization capacity of foundation models with the communication-aware nature of federated learning, FFMs offer the potential to support robust, adaptive, and autonomous operation across physically distributed and intermittently connected platforms.
Rather than viewing FFMs in harsh environments as simply scaling models to the edge, we argue they must be designed as self-sustaining systems that are capable of learning, adapting, and surviving in the face of unreliability and change.

\bibliographystyle{IEEEtran}
\bibliography{reference}

\begin{thebibliography}{10}
\providecommand{\url}[1]{#1}
\csname url@samestyle\endcsname
\providecommand{\newblock}{\relax}
\providecommand{\bibinfo}[2]{#2}
\providecommand{\BIBentrySTDinterwordspacing}{\spaceskip=0pt\relax}
\providecommand{\BIBentryALTinterwordstretchfactor}{4}
\providecommand{\BIBentryALTinterwordspacing}{\spaceskip=\fontdimen2\font plus
\BIBentryALTinterwordstretchfactor\fontdimen3\font minus \fontdimen4\font\relax}
\providecommand{\BIBforeignlanguage}[2]{{%
\expandafter\ifx\csname l@#1\endcsname\relax
\typeout{** WARNING: IEEEtran.bst: No hyphenation pattern has been}%
\typeout{** loaded for the language `#1'. Using the pattern for}%
\typeout{** the default language instead.}%
\else
\language=\csname l@#1\endcsname
\fi
#2}}
\providecommand{\BIBdecl}{\relax}
\BIBdecl

\bibitem{jordan2015machine}
M.~I. Jordan and T.~M. Mitchell, ``Machine learning: Trends, perspectives, and prospects,'' \emph{Science}, vol. 349, no. 6245, pp. 255--260, 2015.

\bibitem{chang2024survey}
Y.~Chang, X.~Wang, J.~Wang, Y.~Wu, L.~Yang, K.~Zhu, H.~Chen, X.~Yi, C.~Wang, Y.~Wang \emph{et~al.}, ``A survey on evaluation of large language models,'' \emph{ACM transactions on intelligent systems and technology}, vol.~15, no.~3, pp. 1--45, 2024.

\bibitem{kairouz2021advances}
P.~Kairouz, H.~B. McMahan, B.~Avent, A.~Bellet, M.~Bennis, A.~N. Bhagoji, K.~Bonawitz, Z.~Charles, G.~Cormode, R.~Cummings \emph{et~al.}, ``Advances and open problems in federated learning,'' \emph{Foundations and trends{\textregistered} in machine learning}, vol.~14, no. 1--2, pp. 1--210, 2021.

\bibitem{zhuang2023foundation}
W.~Zhuang, C.~Chen, and L.~Lyu, ``When foundation model meets federated learning: Motivations, challenges, and future directions,'' \emph{arXiv preprint arXiv:2306.15546}, 2023.

\bibitem{ren2025advances}
C.~Ren, H.~Yu, H.~Peng, X.~Tang, B.~Zhao, L.~Yi, A.~Z. Tan, Y.~Gao, A.~Li, X.~Li \emph{et~al.}, ``Advances and open challenges in federated foundation models,'' \emph{IEEE Communications Surveys \& Tutorials}, 2025.

\bibitem{el2019case}
Z.~El~Khaled and H.~Mcheick, ``Case studies of communications systems during harsh environments: A review of approaches, weaknesses, and limitations to improve quality of service,'' \emph{International journal of distributed sensor networks}, vol.~15, no.~2, p. 1550147719829960, 2019.

\bibitem{feng2023fedmultimodal}
T.~Feng, D.~Bose, T.~Zhang, R.~Hebbar, A.~Ramakrishna, R.~Gupta, M.~Zhang, S.~Avestimehr, and S.~Narayanan, ``Fedmultimodal: A benchmark for multimodal federated learning,'' in \emph{Proceedings of the 29th ACM SIGKDD conference on knowledge discovery and data mining}, 2023, pp. 4035--4045.

\bibitem{li2024multimodal}
C.~Li, Z.~Gan, Z.~Yang, J.~Yang, L.~Li, L.~Wang, and J.~Gao, ``Multimodal foundation models: From specialists to general-purpose assistants,'' 2024.

\bibitem{yarkan2009underground}
S.~Yarkan, S.~Guzelgoz, H.~Arslan, and R.~R. Murphy, ``Underground mine communications: A survey,'' \emph{IEEE Communications Surveys \& Tutorials}, vol.~11, no.~3, pp. 125--142, 2009.

\bibitem{fattah2020survey}
S.~Fattah, A.~Gani, I.~Ahmedy, M.~Y.~I. Idris, and I.~A. Targio~Hashem, ``A survey on underwater wireless sensor networks: Requirements, taxonomy, recent advances, and open research challenges,'' \emph{Sensors}, vol.~20, no.~18, p. 5393, 2020.

\bibitem{kodheli2020satellite}
O.~Kodheli, E.~Lagunas, N.~Maturo, S.~K. Sharma, B.~Shankar, J.~F.~M. Montoya, J.~C.~M. Duncan, D.~Spano, S.~Chatzinotas, S.~Kisseleff \emph{et~al.}, ``Satellite communications in the new space era: A survey and future challenges,'' \emph{IEEE Communications Surveys \& Tutorials}, vol.~23, no.~1, pp. 70--109, 2020.

\bibitem{han2024cooperative}
D.-J. Han, S.~Hosseinalipour, D.~J. Love, M.~Chiang, and C.~G. Brinton, ``Cooperative federated learning over ground-to-satellite integrated networks: Joint local computation and data offloading,'' \emph{IEEE Journal on Selected Areas in Communications}, vol.~42, no.~5, pp. 1080--1096, 2024.

\bibitem{wang2024fedmeki}
J.~Wang, X.~Wang, L.~Lyu, J.~Chen, and F.~Ma, ``Fedmeki: A benchmark for scaling medical foundation models via federated knowledge injection,'' \emph{Advances in Neural Information Processing Systems}, vol.~37, pp. 1082--1116, 2024.

\bibitem{kikaki2024detecting}
K.~Kikaki, I.~Kakogeorgiou, I.~Hoteit, and K.~Karantzalos, ``Detecting marine pollutants and sea surface features with deep learning in sentinel-2 imagery,'' \emph{ISPRS Journal of Photogrammetry and Remote Sensing}, vol. 210, pp. 39--54, 2024.

\bibitem{islam2020semantic}
M.~J. Islam, C.~Edge, Y.~Xiao, P.~Luo, M.~Mehtaz, C.~Morse, S.~S. Enan, and J.~Sattar, ``Semantic segmentation of underwater imagery: Dataset and benchmark,'' in \emph{2020 IEEE/RSJ international conference on intelligent robots and systems (IROS)}.\hskip 1em plus 0.5em minus 0.4em\relax IEEE, 2020, pp. 1769--1776.

\end{thebibliography}

\begin{IEEEbiographynophoto}{Evan Chen (M’23)}
is a PhD student of Electrical and Computer Engineering (ECE) at Purdue University.
\end{IEEEbiographynophoto}
\begin{IEEEbiographynophoto}{Seyyedali Hosseinalipour (SM’25)}
is an Assistant Professor of Electrical Engineering (EE) at University at Buffalo (SUNY).
\end{IEEEbiographynophoto}
\begin{IEEEbiographynophoto}{Christopher G. Brinton (SM’20)}
is the Elmore Associate Professor of ECE at Purdue University.
\end{IEEEbiographynophoto}
\begin{IEEEbiographynophoto}{David Love (F’15)}
is the Nick Trbovich Professor of ECE at Purdue University.
\end{IEEEbiographynophoto}

\end{document}